\documentclass[a4paper,10pt]{article}
\usepackage[utf8x]{inputenc}
\usepackage{amsmath}

\title{On the New Dimer $\lambda_d$ $x$-Expansion, \\ Triangular and Hexagonal Lattices too}
\author{Paul Federbush \\
Department of Mathematics \\
University of Michigan \\
Ann Arbor, MI 48109-1043 \\
(pfed@umich.edu)}

\begin{document}

\maketitle

\begin{abstract}
In recent work the author presented a formal expansion for $\lambda_d$ associated to the dimer problem on a $d$-dimensional rectangular lattice.  Expressed in terms of $d$ it yielded a presumed asymptotic expansion for $\lambda_d$ in inverse powers of $d$.  We also considered an expansion in powers of $x$, a formal variable ultimately set equal to $1$.  We believe this series has better asymptotic properties than the expansion in inverse powers of $d$.  We discuss this, and apply the same method to the two-dimensional triangular and hexagonal lattices.  Viewed as a test of the $x$-expansion the results on those two lattices are satisfactory, if not thoroughly convincing.
\end{abstract}

The formalism we will be working within is presented in \cite{Federbush}, a much clearer exposition than in the earlier reports on which \cite{Federbush} is based.  The main result there is the formal asymptotic expansion for $\lambda_d$ (called by some the entropy per site)
\begin{align}
\label{lambda_d}
\lambda_d \sim \frac{1}{2} \ln \left( 2d \right) - \frac{1}{2} + \frac{1}{8} \frac{1}{d} + \frac{5}{96} \frac{1}{d^2} + \frac{5}{64} \frac{1}{d^3} \cdots
\end{align}
Here $d$ is the dimension of the rectangular lattice on which one is studying the dimer problem.  However we prefer now to define $d$, instead, as one half the number of edges entering each vertex.

The key quantities determining the expansion are the $\bar{J}_n$, $n=1,2, \dots$, certain cluster expansion kernels.  For the rectangular lattices the $\bar{J}_n$, $n=1, \dots, 6$, are given in (31) - (36) of \cite{Federbush}.  We now present the result of computer computations for some of these $\bar{J}_n$ for the triangular and hexagonal lattices.  For the triangular lattice one has
\begin{align}
\bar{J}_1 &= 0 \label{J1-tri} \\
\bar{J}_2 &= \frac{1}{24} \label{J2-tri} \\
\bar{J}_3 &= 0 \label{J3-tri} \\
\bar{J}_4 &= - \frac{31}{1728} \label{J4-tri} \\
\bar{J}_5 &= - \frac{13}{6480}. \label{J5-tri}
\end{align}
And for the hexagonal lattice
\begin{align}
\bar{J}_1 &= 0 \label{J1} \\
\bar{J}_2 &= \frac{1}{12} \label{J2-hex} \\
\bar{J}_3 &= \frac{1}{27} \label{J3-hex} \\
\bar{J}_4 &= - \frac{7}{216} \label{J4-hex} \\
\bar{J}_5 &= - \frac{23}{405} \label{J5-hex} \\
\bar{J}_6 &= \frac{23787}{8748}. \label{J6-hex}
\end{align}

The formal expression for $\lambda_d$, in terms of the $\bar{J}_n$ is given as
\begin{align}
\label{lambda}
\lambda_d = \frac{1}{2} \ln \left(2d \right) - \frac{1}{2} + \frac{1}{N} \ln Z^*
\end{align}
where $Z^*$ is detailed in (28) - (30) of \cite{Federbush}.  If one expresses $Z^*$ in terms of $d$, and expands in powers of ${\raise0.5ex\hbox{$\scriptstyle 1$}} \kern-0.1em/\kern-0.15em \lower0.25ex\hbox{$\scriptstyle d$}$ one gets (\ref{lambda_d}).

Alternatively we introduce a parameter $x$, replacing $\bar{J}_n$ by $x^{n-1} \bar{J}_n$.  We then express $\frac{1}{N} \ln Z^{*}$ as a function of $x$ and the $\bar{J}_n$ and expand in powers of $x$.  This leads to a formal expression for $\lambda_d$ as
\begin{align}
\label{lambda-x}
\lambda = & \frac{1}{2} \ln \left(2d \right) - \frac{1}{2} + \left( \bar{J}_2 \right) x + \left( \bar{J}_3 + 4 \bar{J}_2 ^2 \right) x^2 + \left( \frac{112}{3} \bar{J}_2 ^3 + 12 \bar{J}_2 \bar{J}_3 + \bar{J}_4 \right) x^3 \notag \\ 
& + \left( \bar{J}_5 + 16 \bar{J}_2 \bar{J}_4 + 192 \bar{J}_2 ^2 \bar{J}_3 + 9 \bar{J}_3 ^2 + 480 \bar{J}_2 ^4 \right) x^4 \notag \\
& + \left( \frac{36608}{5} \bar{J}_2 ^5 + \bar{J}_6 + 20 \bar{J}_2 \bar{J}_5 + 24 \bar{J}_3 \bar{J}_4 + 288 \bar{J}_4 \bar{J}_2 ^2 + 324 \bar{J}_3 ^2 \bar{J}_2 + 3520 \bar{J}_2 ^3 \bar{J}_3 \right) x^5 \notag \\
& + \cdots
\end{align}
with $x$ set equal to $1$ herein.  We use $\lambda$ instead of $\lambda_d$ since we may apply the same expression for other lattices as the triangular and hexagonal.  This is the $x$-expansion for $\lambda$.

If we substitute into equation (\ref{lambda-x}) the values of the $\bar{J}_n$ given in (31) - (36) of \cite{Federbush} we obtain another expression for $\lambda_d$.  Keeping only the terms in it with inverse powers of $d$ less than or equal to $3$ we recover (\ref{lambda_d}).  The expression obtained from (\ref{lambda_d}) and that from (\ref{lambda-x}) are each as far as one can go with the information in the $\bar{J}_n$ for $n \le 6$.

We let $B_r$ be obtained from (\ref{lambda-x}) by keeping powers of $x$ through $x^r$ (and again then setting $x=1$).  Thus the expression on the right side of equation (\ref{lambda-x}), before the dots, is $B_5$, and
\begin{align}
B_0 = \frac{1}{2} \ln \left( 2d \right) - \frac{1}{2}.
\end{align}
We assemble a table of $B$ values for the rectangular lattices of dimensions $1$, $2$, and $3$, columns $1$, $2$, and $3$ below; and for the hexagonal and triangular lattices in columns $4$ and $5$ respectively.  
\begin{center}
  \begin{tabular}{ l | r | r | r | r | r | }
   \ & 1 & 2 & 3 & 4 & 5 \\ \hline
    $B_0$ & -.1534 & .1931 & .3959 & .0493 & .3959 \\
    $B_1$ & -.0284 & .2556 & .4375 & .1326 & .4375 \\
    $B_2$ & .1174 & .2921 & .4538 & .1975 & .4445 \\
    $B_3$ & .2684 & .2993 & .4524 & .2237 & .4293 \\
    $B_4$ & .4419 & .2906 & .4468 & .2086 & .4167 \\
    $B_5$ & .7096 & .2814 & .4445 & 2.88 & $-$ \\
  \end{tabular}
\end{center}
The values of $\lambda$ for each of these are known as follows:
\begin{align}
\lambda_1 &= 0 \label{lam1} \\
\lambda_2 &= .29156 \dots \label{lam2} \\
.440075 \le \lambda_3 & \le .457547 \label{lam3} \\
\lambda_{\rm hex} &= .1691 \dots \label{lamhex} \\
\lambda_{\rm tri} &= .4286 \dots \label{lamtri}
\end{align}
Equation (\ref{lam1}) is an easy deduction.  References for (\ref{lam2}) and (\ref{lam3}) are in \cite{Federbush}, for (\ref{lamhex}) see \cite{NWW}, and for (\ref{lamtri}) see \cite{NGS}.  It is too bad we do not have many more known lattices to test our ideas against.  Would that $\lambda_3$ and $\lambda_4$ were known say.

Once we hoped $\lim_{n \to \infty} B_n = \lambda$, now we believe instead that the $x$-expansion behaves as an asymptotic series.  The prototype asymptotic series is a sum of terms that get smaller and smaller to some point and then quickly grow.  In favorable situations a ``good'' sum for the series is obtained by keeping only the terms till they start growing (we will actually keep the smallest term times ${\raise0.5ex\hbox{$\scriptstyle 1$}} \kern-0.1em/\kern-0.15em \lower0.25ex\hbox{$\scriptstyle 2$}$, and throw out the succeeding terms).  Thus we take as value of the sum $\frac{1}{2} \left( B_k + B_{k+1} \right)$ where $k$ is such that $\left| B_{k+1}- B_k \right|$ is minimum in value.  We take the size of the smallest term, $\left| B_{k+1}- B_k \right|$, as an estimate of the error of our approximation.  Applying this procedure to obtain a value of $\lambda$ from the $x$-expansion, we get in each of these five cases:
\begin{align}
\lambda_1 & \cong .03 \pm .13 \label{lamx1} \\
\lambda_2 & \cong .296 \pm .007 \label{lamx2} \\
\lambda_3 & \cong .453 \pm .001 \label{lamx3} \\
\lambda_{\rm hex} & \cong .216 \pm .015 \label{lamxhex} \\
\lambda_{\rm tri} & \cong .441 \pm .007 \label{lamxtri}
\end{align}
We believe, for $\lambda_d$, the results of this procedure will be increasingly strikingly accurate as $d \to \infty$, and even for $d=2$ and $3$, as here, are impressive.  We take (\ref{lamx3}) as a good prediction for $\lambda_3$.  For $\lambda_{\rm hex}$ and $\lambda_{\rm tri}$ the listed error estimates in (\ref{lamxhex}) and (\ref{lamxtri}) must be tripled and doubled respectively.  We note eq. (\ref{lambda_d}) gives for $\lambda_1$, $\lambda_2$,  $\lambda_3$ the values $.102$, $.278$, $.446$.  Both eq. (\ref{lambda_d}) and the $x$-expansion procedure are pleasingly accurate at $\lambda_3$.

The expansions for $\lambda$ are for dimension going to infinity.  We should be pleased they yield interesting results even in two dimensions.  Seeking a better way to extract $\lambda$ from the sequence of $\bar{J}_n$  may be like searching for the pot of gold at the end of the rainbow.
\pagebreak


\begin{thebibliography}{}

\bibitem{Federbush} Paul Federbush, Phys. Lett. A \textbf{374} (2009) 131.

\bibitem{NWW} J F Nagle, Phys. Rev. Lett. \textbf{34} (1975) 1150 \\
G H Wannier, Phys. Rev. \textbf{79} (1950) 357 \\ 
F Y Wu, Phys. Rev. \textbf{168} (1968) 539 \\
P W Kasteleyn, J. Math. Phys. \textbf{4} (1963) 287 \\
Veit Elser, J. Phys. A: Math. Gen. \textbf{17} (1984) 1509.

\bibitem{NGS} J F Nagle, Phys. Rev. \textbf{152} (1966) 190 \\
D S Gaunt, Phys. Rev. \textbf{179} (1969) 174 \\
S Samuel, J. Math. Phys. \textbf{21} (1980) 2806 \\
F Mila, Phys. Rev. Lett. \textbf{81} (1998) 2356 \\
W F Wolff and J Zittartz, Z. Phys. B\textbf{49} (1982) 139 \\
P Fendley, R Moessner, S L Sondhi, Phys. Rev. B \textbf{66} (2002) 214513.

\end{thebibliography}
\end{document}